\documentclass[sigconf, screen, nonacm]{acmart} %

\newif\ifanonymous
\anonymousfalse  %

\ifanonymous
  \newcommand{\ToolEval}{[EvalTool]}      %
  \newcommand{\ToolGen}{[GenTool]}        %
  \newcommand{\selfcite}[1][]{[anonymous]}  %
\else
  \newcommand{\ToolEval}{JudgeGPT}
  \newcommand{\ToolGen}{RogueGPT}

  \newcommand{\selfcite}[1][]{\cite{loth_blessing_2024,loth2026eroding,loth2026roguegpt,loth2026verification,loth2026originlens}}
\fi

\copyrightyear{2026}
\acmYear{2026}
\setcopyright{cc}
\setcctype{by}
\acmConference[WWW Companion '26]{Companion Proceedings of the ACM Web Conference 2026}{April 13--17, 2026}{Dubai, United Arab Emirates.}
\acmBooktitle{Companion Proceedings of the ACM Web Conference 2026 (WWW Companion '26), April 13--17, 2026, Dubai, United Arab Emirates}
\acmPrice{}
\acmDOI{10.1145/3774905.3795471}
\acmISBN{979-8-4007-2308-7/2026/04}

\usepackage{algorithmic}
\usepackage{graphicx}
\usepackage{subfigure}
\usepackage{textcomp}

\usepackage[hyphenbreaks]{breakurl}
\usepackage{xurl}
\def\BibTeX{{\rm B\kern-.05em{\sc i\kern-.025em b}\kern-.08em
    T\kern-.1667em\lower.7ex\hbox{E}\kern-.125emX}}
\usepackage{cleveref}
\usepackage{booktabs}
    
\usepackage{xcolor}

 \usepackage{pifont}%

\usepackage{listings}
\usepackage{fancyvrb}
\usepackage{framed}
\usepackage{color}
\usepackage{pgfplots}
\pgfplotsset{compat=1.18}
\usetikzlibrary{positioning, decorations.pathreplacing, arrows.meta, calc}
\usepackage{tabularx}
\definecolor{light-gray}{gray}{0.95}
\hypersetup{
    breaklinks=true,   %
    colorlinks=true    %
}

\usepackage{fancyhdr}

\begin{document}

\title[Industrialized Deception via GenAI]{Industrialized Deception: The Collateral Effects of LLM-Generated Misinformation on Digital Ecosystems}
\titlenote{This research is funded by the European Union project CyberSecDome under the grant agreement 101120779. Published in WWW Companion '26, pp.~295--302. DOI: \href{https://doi.org/10.1145/3774905.3795471}{10.1145/3774905.3795471}.}

\author{Alexander Loth}
\email{alexander.loth@stud.fra-uas.de}
\orcid{0009-0003-9327-6865}
\affiliation{%
  \institution{Frankfurt University of Applied Sciences}
  \city{Frankfurt am Main}
  \country{Germany}
}

\author{Martin Kappes}
\email{kappes@fra-uas.de}
\orcid{0000-0002-8768-8359}
\affiliation{%
  \institution{Frankfurt University of Applied Sciences}
  \city{Frankfurt am Main}
  \country{Germany}
}

\author{Marc-Oliver Pahl}
\email{marc-oliver.pahl@imt-atlantique.fr}
\orcid{0000-0001-5241-3809}
\affiliation{%
  \institution{IMT Atlantique, UMR IRISA, Chaire Cyber CNI}
  \city{Rennes}
  \country{France}
}

\begin{abstract}
Generative AI and misinformation research have evolved since our 2024 survey.
This paper presents an updated perspective, transitioning from literature review to practical countermeasures.
We report on changes in the threat landscape, including improved AI-generated content through Large Language Models (LLMs) and multimodal systems -- addressing the adverse impacts of AI on information quality.

This work focuses on our practical contributions: \textit{JudgeGPT}, a platform to evaluate human perception of AI-generated news, and \textit{RogueGPT}, a controlled stimulus generation engine for research.
Together, these tools form an experimental pipeline for studying how humans perceive and detect AI-generated misinformation.

Our findings show that detection capabilities have improved, but competition between generation and detection continues.
We discuss mitigation strategies such as LLM-based detection, inoculation approaches, and the dual-use nature of generative AI.
\end{abstract}



\renewcommand{\shortauthors}{Alexander Loth et al.}

\maketitle

\pagestyle{fancy}
\fancyhf{}
\fancyhead[L]{\textit{A.\ Loth, M.\ Kappes, M.-O.\ Pahl}}
\fancyhead[R]{\textit{Published in WWW Companion '26, pp.\ 295--302}}
\fancyfoot[C]{\thepage}
\renewcommand{\headrulewidth}{0pt}

\section{Introduction}
\label{introduction}

Generative AI has changed how information---and misinformation---spreads online.
The ability to generate convincing text at scale has enabled what can be termed \emph{industrialized deception}: the automated production of misleading content affecting digital ecosystems.
These digital ecosystems---comprising interconnected networks of platforms, users, algorithms, and content---have become the primary infrastructure through which information flows in modern society~\cite{fastowski_knowledge_drift_2024}.
The health of these ecosystems depends on the trustworthiness of information circulating within them; when misinformation proliferates, it erodes trust not only in specific content but in the information infrastructure itself.
In our 2024 survey~\cite{loth_blessing_2024}, we examined the interplay between Generative AI and Fake News, covering enabling technologies, content creation, detection methods, and deepfake threats.
Since then, the field has evolved in ways that warrant renewed examination.

Large Language Models (LLMs) have improved considerably, intensifying the competition between generation and detection of synthetic content~\cite{bubeck_sparks_2023,sallami_deception_2024}.
Researchers have explored the dual-use nature of LLMs---their capacity to both generate and detect misinformation~\cite{sallami_exploring_2025,herder_preventing_2025}---with societal implications extending to privacy, manipulation, and trust erosion~\cite{aimeur_privacy_2025}.

This paper makes three key contributions.
First, we provide an updated perspective on how the Generative AI and Fake News research domain has evolved since 2024, highlighting new challenges including the emergence of multimodal misinformation---where text, images, audio, and video are combined to create more convincing deceptive content~\cite{ai_paradigm_2026,hussain_generalized_2025}---and the shift toward agentic AI systems capable of autonomous content generation and dissemination, which motivates moving beyond content-level detection toward behavioral-level analysis of coordinated inauthentic behavior~\cite{tseng2026agentic,acmetpc2025systemic}.
Second, we present our methodological contributions: the open-source tools \textit{\ToolEval}\footnote{\url{https://github.com/aloth/JudgeGPT}}~\cite{loth2026eroding} and \textit{\ToolGen}\footnote{\url{https://github.com/aloth/RogueGPT}}, which together form an experimental pipeline for studying human perception of AI-generated news.
\ToolEval{} serves as an empirical data collection platform where participants evaluate news fragment authenticity, while \ToolGen{} provides controlled stimulus generation for research purposes.
Our longitudinal expert perception survey reveals that large-scale text generation poses systemic risks of ``epistemic fragmentation'' and ``synthetic consensus''---risks now formalized in Ferrara's ``Generative AI Paradox'' framework~\cite{Ferrara2026Paradox}---while experts express skepticism toward purely technical detection tools, preferring provenance standards and regulatory frameworks aligned with emerging ``epistemic security'' objectives~\cite{loth2026verification,IPIE2025Epistemic}.
Third, we discuss emerging mitigation strategies and the role of AI itself in combating misinformation, including inoculation theory~\cite{lewandowsky_countering_2021} and LLM-based detection approaches~\cite{herder_preventing_2025}.

Our work contributes to efforts addressing AI's adverse impacts and collateral effects by examining both threats and countermeasures~\cite{tommasel_countering_2025,godoy_moral_2024}.
The dual nature of Generative AI---as both a tool for creating deceptive content and for detecting it---warrants continued research as these technologies become more widely deployed.

This paper is structured as follows:
\Cref{domain} introduces key concepts and recent developments.
\Cref{overview} reviews relevant research since 2024.
\Cref{findings} presents our methodological contributions.
\Cref{state} synthesizes findings and discusses mitigation strategies.
\Cref{conclusion} presents conclusions and future directions.
\section{The Domain}
\label{domain}

The term ``Generative Artificial Intelligence (AI)'' refers to AI technologies designed to produce new content. 
This content includes text, images, audio, and other media forms, often resembling human-generated output.
Models like GPT-4 can produce text that evaluators struggle to distinguish from human writing~\cite{biever_chatgpt_2023}.

Machine learning models such as Generative Adversarial Networks (GANs), transformers, and variational autoencoders enable this capability. 
Trained on large datasets, these models learn to generate new instances that reflect patterns in their training data~\cite{cao_comprehensive_2023,gozalo-brizuela_chatgpt_2023}.

\subsection{Evolution Since 2024}

Since our 2024 survey~\cite{loth_blessing_2024}, several developments have changed the Generative AI and Fake News research area:

\textbf{LLMs and Multimodal Systems.} Models like GPT-4o, Claude 3.5, and Gemini 1.5 have improved at generating coherent content across multiple modalities~\cite{ferrara_genai_2024}.
Reasoning models (e.g., OpenAI o1) and small language models for edge deployment have widened access to generation capabilities~\cite{kumar2025peeping}.
The emergence of large vision-language models (LVLMs) has changed both generation and detection of multimodal content~\cite{ai_paradigm_2026}.

\textbf{Multimodal Misinformation Challenges.} The combination of text, images, audio, and video in misinformation poses detection challenges that exceed single-modality approaches.
Out-of-context misinformation---where authentic content is paired with misleading narratives---has emerged as a common form requiring cross-modal semantic analysis~\cite{li_multicaption_2026}.
Recent advances include agentic frameworks that use web-grounded reasoning for verification~\cite{shopnil_mirage_2025} and tool-augmented detection agents~\cite{cui_t2agent_2026}.

\textbf{Dual-Use Nature of LLMs.} Recent research has explored how the same LLMs that can generate misinformation can also be used for detection~\cite{sallami_deception_2024,herder_preventing_2025}.
Sallami et al. demonstrate that LLMs exhibit both creative capabilities for generating convincing fake content and analytical capabilities for identifying it~\cite{sallami_deception_2024}.

\textbf{Bias and Fairness Concerns.} The research community has increasingly focused on biases embedded in AI detection systems, including gender bias in Fake News detection~\cite{sallami_gender_2024} and the need for fairness frameworks~\cite{sallami_fairframe_2025}.

\textbf{Agentic AI and the Operationalization of Influence.} A notable development since 2024 is the emergence of agentic AI as a vehicle for industrialized deception.
The threat model has shifted from human actors leveraging GenAI tools to autonomous agents capable of independent reasoning, planning, and execution~\cite{gartner2025agentic}.
Tseng et al. (2026) provide evidence of multi-agent pipelines systematizing Foreign Information Manipulation and Interference (FIMI), with specialized agentic components mapping behaviors to the DISARM framework~\cite{tseng2026agentic}.
This represents a shift from a ``content abundance'' problem to a ``coordination abundance'' problem---the constraint on disinformation campaigns is no longer human labor but compute~\cite{acmetpc2025systemic}.
Autonomous agents can perceive information environments, reason about psychological triggers, generate tailored multimodal content, and refine strategies based on real-time engagement metrics without human intervention.

\textbf{Beyond Detection to Prevention.} Research focus has shifted from detection toward prevention strategies, including inoculation approaches~\cite{lewandowsky_countering_2021} and ``prebunking'' techniques~\cite{sallami_exploring_2025}.

As Generative AI improves, it both enables synthetic content creation and provides tools for detection.
This dual-use nature motivates work on content authenticity verification.
Cryptographic provenance standards such as C2PA offer an alternative to detection by establishing verifiable chains of content origin; our Origin Lens framework implements privacy-preserving on-device verification using a defense-in-depth approach~\cite{loth2021decisively,loth2026originlens}.

\subsection{Structural Overview}

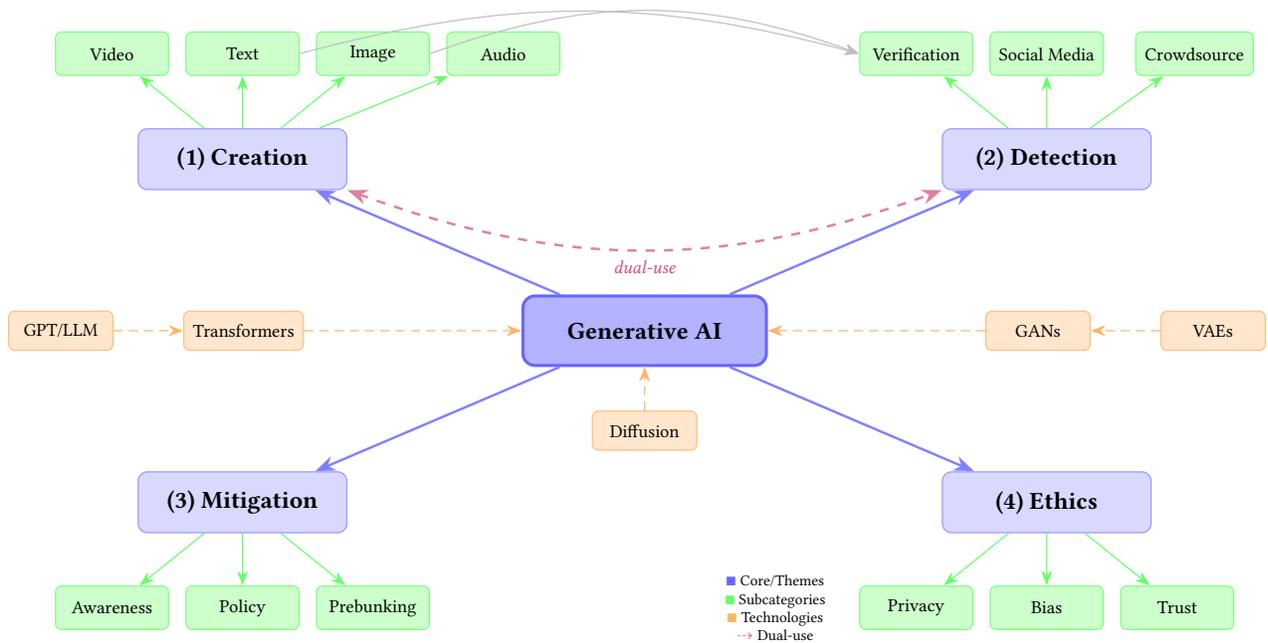
\begin{figure*}[ht!]
\centering
\resizebox{0.95\textwidth}{!}
{
\begin{tikzpicture}[
  >=Stealth,
  node distance=0.8cm and 1.2cm,
  every node/.style={font=\small},
  core/.style={rectangle, rounded corners=4pt, fill=blue!30, draw=blue!60, line width=1pt, minimum width=2.8cm, minimum height=0.8cm, font=\small\bfseries},
  theme/.style={rectangle, rounded corners=3pt, fill=blue!15, draw=blue!40, minimum width=2.4cm, minimum height=0.7cm, align=center, font=\small\bfseries},
  subcat/.style={rectangle, rounded corners=2pt, fill=green!20, draw=green!50, minimum width=1.3cm, minimum height=0.5cm, align=center, font=\scriptsize},
  tech/.style={rectangle, rounded corners=2pt, fill=orange!20, draw=orange!50, minimum width=1.2cm, minimum height=0.45cm, align=center, font=\scriptsize},
  mainedge/.style={->, thick, blue!50},
  subedge/.style={->, thin, green!60},
  techedge/.style={->, densely dashed, orange!60},
  dualedge/.style={<->, thick, purple!50, dashed}
]

\node[core] (genai) {Generative AI};

\node[theme] (creation) [above left=1.2cm and 2cm of genai] {(1) Creation};
\node[theme] (detection) [above right=1.2cm and 2cm of genai] {(2) Detection};
\node[theme] (mitigation) [below left=1.2cm and 2cm of genai] {(3) Mitigation};
\node[theme] (ethics) [below right=1.2cm and 2cm of genai] {(4) Ethics};

\node[subcat] (text) [above=0.6cm of creation, xshift=0cm] {Text};
\node[subcat] (image) [above=0.6cm of creation, xshift=1.5cm] {Image};
\node[subcat] (audio) [above=0.6cm of creation, xshift=3.0cm] {Audio};
\node[subcat] (video) [above=0.6cm of creation, xshift=-1.5cm] {Video};

\node[subcat] (verify) [above=0.6cm of detection, xshift=-1.5cm] {Verification};
\node[subcat] (social) [above=0.6cm of detection, xshift=0cm] {Social Media};
\node[subcat] (crowd) [above=0.6cm of detection, xshift=1.7cm] {Crowdsource};

\node[subcat] (aware) [below=0.6cm of mitigation, xshift=-1.5cm] {Awareness};
\node[subcat] (policy) [below=0.6cm of mitigation, xshift=0cm] {Policy};
\node[subcat] (prebunk) [below=0.6cm of mitigation, xshift=1.5cm] {Prebunking};

\node[subcat] (privacy) [below=0.6cm of ethics, xshift=-1.5cm] {Privacy};
\node[subcat] (bias) [below=0.6cm of ethics, xshift=0cm] {Bias};
\node[subcat] (trust) [below=0.6cm of ethics, xshift=1.5cm] {Trust};

\node[tech] (trans) [left=2.5cm of genai] {Transformers};
\node[tech] (gpt) [left=0.8cm of trans] {GPT/LLM};
\node[tech] (gan) [right=2.5cm of genai] {GANs};
\node[tech] (vae) [right=0.8cm of gan] {VAEs};
\node[tech] (diff) [below=0.5cm of genai] {Diffusion};

\draw[mainedge] (genai) -- (creation);
\draw[mainedge] (genai) -- (detection);
\draw[mainedge] (genai) -- (mitigation);
\draw[mainedge] (genai) -- (ethics);

\draw[subedge] (creation) -- (text);
\draw[subedge] (creation) -- (image);
\draw[subedge] (creation) -- (audio);
\draw[subedge] (creation) -- (video);

\draw[subedge] (detection) -- (verify);
\draw[subedge] (detection) -- (social);
\draw[subedge] (detection) -- (crowd);

\draw[subedge] (mitigation) -- (aware);
\draw[subedge] (mitigation) -- (policy);
\draw[subedge] (mitigation) -- (prebunk);

\draw[subedge] (ethics) -- (privacy);
\draw[subedge] (ethics) -- (bias);
\draw[subedge] (ethics) -- (trust);

\draw[techedge] (trans) -- (genai);
\draw[techedge] (gpt) -- (trans);
\draw[techedge] (gan) -- (genai);
\draw[techedge] (vae) -- (gan);
\draw[techedge] (diff) -- (genai);

\draw[dualedge, bend right=20] (creation.south east) to node[below, font=\scriptsize\itshape, text=purple!70] {dual-use} (detection.south west);

\draw[thin, gray!50, ->] (text.east) to[bend left=15] (verify.west);
\draw[thin, gray!50, ->] (image.east) to[bend left=20] (verify.west);

\node[font=\tiny, align=center] at ($(prebunk.east)!0.5!(trust.west)$) {
\textcolor{blue!60}{$\blacksquare$} Core/Themes\\
\textcolor{green!60}{$\blacksquare$} Subcategories\\  
\textcolor{orange!60}{$\blacksquare$} Technologies\\
\textcolor{purple!60}{$\dashrightarrow$} Dual-use
};

\end{tikzpicture}
}
\caption{Structural overview of Generative AI's impact on Fake News. The dual-use nature (purple dashed arrow) illustrates how the same technologies enable both creation and detection of synthetic content.}
\Description{Diagram: A central \emph{Generative AI} node connects to four theme nodes (Creation, Detection, Mitigation, Ethics), each with subnodes (e.g., Text/Image/Audio/Video; Verification/Social Media/Crowdsource; Awareness/Policy/Prebunking; Privacy/Bias/Trust). Additional technology nodes (Transformers, GPT/LLM, GANs, VAEs, Diffusion) are linked; a dashed bidirectional arrow highlights the dual-use relationship between Creation and Detection.}
\label{fig:generative-ai-fake-news-structure}
\end{figure*}

Figure~\ref{fig:generative-ai-fake-news-structure} illustrates the domain structure.
At the core, Generative AI branches into two principal areas: creation and detection of Fake News.
The Creation aspect encompasses Text Generation, Image Synthesis, Audio Generation, and Video Generation---representing the diverse capabilities to produce content indistinguishable from human-created material.
The Detection branch addresses Content Verification, Social Media Analysis, and Crowd Sourcing for identifying synthetic content.

Adjacent to these themes are mitigation strategies (public awareness, regulatory policies) and ethical considerations (privacy concerns, bias and fairness)~\cite{aimeur_privacy_2025,knijnenburg_transparency_2024}.
Connecting these nodes are enabling technologies: Autoencoders, GANs, Transformers, GPTs, and VAEs.

\subsection{Digital Ecosystems and Information Integrity}

Digital ecosystems comprise interconnected networks of platforms, users, algorithms, and content that collectively shape how information flows through society.
These ecosystems include social media platforms, search engines, news aggregators, messaging applications, and recommendation systems---each influencing what content users encounter and share~\cite{ferrara_genai_2024}.
The health of digital ecosystems depends on multiple factors: the trustworthiness of information sources, the transparency of algorithmic curation, and the resilience of users to manipulation~\cite{fastowski_knowledge_drift_2024}.

\textbf{From Fake News to Synthetic Reality.}
Ferrara (2026) argues that the prevailing focus on ``deepfakes'' or ``misinformation'' misses a broader socio-technical shift: the creation of \emph{Synthetic Realities}~\cite{Ferrara2026Paradox}.
This framework formalizes the threat as a layered stack: (1) \emph{Synthetic Content}---the raw text, image, audio, or video artifacts; (2) \emph{Synthetic Identity}---the fabrication of coherent personas that persist over time; (3) \emph{Synthetic Interaction}---the simulation of social presence, engagement, and relationship-building; and (4) \emph{Synthetic Institutions}---the manufacture of consensus through coordinated networks of fake outlets and organizations.
This final layer is particularly relevant to ``Industrialized Deception,'' as it implies the automation of credibility itself, not just content.

\textbf{The Generative AI Paradox.}
Ferrara's ``Generative AI Paradox'' posits that as synthetic media becomes ubiquitous and indistinguishable from authentic content, societies will rationally discount \emph{all} digital evidence.
The cost of verification becomes prohibitively high compared to the cost of generation, leading to a market failure in the information ecosystem~\cite{Ferrara2026Paradox}.
Trust is not merely eroded; it is rendered economically irrational.
This aligns with the concept of ``Epistemic Security'' highlighted in recent policy discussions, where the goal of defense shifts from ``correcting false information'' (which assumes a functioning marketplace of ideas) to ``securing the conditions for knowledge creation'' (which acknowledges that the marketplace itself is flooded)~\cite{IPIE2025Epistemic}.

Generative AI poses systemic risks to these ecosystems through several interconnected mechanisms.
First, LLMs enable the production of misleading content at high \emph{scale and speed}, overwhelming traditional fact-checking and moderation systems~\cite{maurya_simulating_2025}.
Second, synthetic content tailored to specific audiences creates \emph{epistemic fragmentation}---information bubbles with incompatible worldviews that fragment shared understanding, a key concern identified in expert surveys~\cite{loth2026verification,IPIE2025Epistemic}.
Third, coordinated deployment of AI-generated content manufactures \emph{synthetic consensus}, exploiting the Synthetic Institutions layer of Ferrara's stack to manipulate perceptions of public opinion~\cite{Ferrara2026Paradox}.
Finally, as users become aware of AI-generated content, skepticism extends to authentic content, embodying the Generative AI Paradox where rational actors discount all digital evidence---a phenomenon we term \emph{trust erosion}.

Platform algorithms amplify these effects by optimizing for engagement metrics that often favor sensational or emotionally charged content---characteristics that AI can readily generate~\cite{herder_junkfood_2024,bradshaw_industrialized_2021}.
Understanding these ecosystem dynamics is useful for developing mitigation strategies that address not just individual pieces of misinformation but the structural conditions enabling their spread.

\subsection{Functioning of Generative AI for Fake News Generation}

Generative AI models synthesize new data by learning from existing datasets. 
They function through deep learning architectures such as GANs, VAEs, and Transformers. 
GANs pit two neural networks against each other to produce new, synthetic instances of data.
GANs generate realistic images and videos to accompany synthetic Fake News stories.

Transformers utilize attention mechanisms to generate coherent sequences of text\cite{vaswani_attention_2023}.
Transformers, like GPT models, are trained on vast corpora of text.
Transformers are able to produce all kind of text, including Fake News\cite{sandrini_generative_2023, ferrara_genai_2024}.

\subsection{Technical Background}
This section provides an overview of the key technologies and concepts foundational to understanding the intersection of Generative AI and Fake News. It introduces the essential definitions and methodologies employed in the survey.

\subsubsection{Generative Artificial Intelligence}
Generative AI refers to a subset of AI technologies designed to create content that mimics real-world data. These models learn to generate new data samples that are indistinguishable from authentic datasets.

\subsubsection{Generative Adversarial Networks (GANs)}
GANs consist of two neural networks, the generator and the discriminator, which are trained simultaneously through adversarial processes. The generator creates data samples aimed at fooling the discriminator, while the discriminator evaluates them against real data, improving both models iteratively\cite{goodfellow_generative_2014}.

\subsubsection{Variational Autoencoders (VAEs)}
VAEs are generative models that use a probabilistic approach to produce data. They learn to encode input data into a latent space and reconstruct it back, ensuring that generated samples adhere to the probability distribution of the input data\cite{kingma_auto-encoding_2022}.

\subsubsection{Transformer Models}
Transformers are a type of neural network architecture designed for processing sequential data, particularly text.
They rely on self-attention mechanisms to weigh the significance of different parts of the input data\cite{vaswani_attention_2023}.
A recent breakthrough in this area is the development of 1-bit Large Language Models (LLMs), such as those introduced by Ma et al. (2024)\cite{ma_era_2024}, which achieve comparable performance to full-precision models with significantly reduced computational costs.

\paragraph{BERT}
Bidirectional Encoder Representations from Transformers (BERT) is a model designed to pre-train deep bidirectional representations from unlabeled text by jointly conditioning on both left and right context in all layers\cite{devlin_bert_2018}.

\paragraph{GPT}
Generative Pre-trained Transformer (GPT) models generate coherent text based on a given prompt.
These models can perform various language tasks without task-specific training~\cite{radford_improving_2018}.

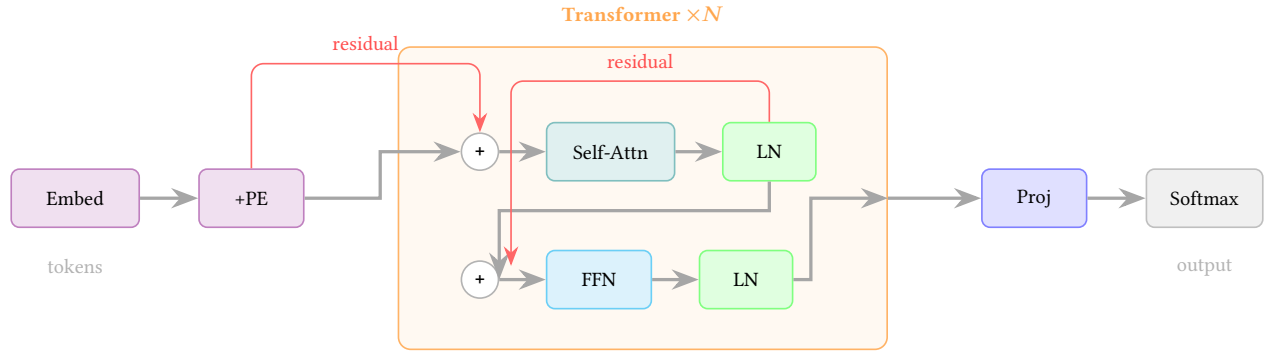
\begin{figure*}[ht!]
\centering
\resizebox{0.95\textwidth}{!}{
\begin{tikzpicture}[
  >=Stealth,
  node distance=0.4cm,
  comp/.style={draw, rectangle, rounded corners=2pt, minimum height=0.5cm, minimum width=1.1cm, font=\tiny, fill=#1!12, draw=#1!50},
  comp/.default=blue,
  arr/.style={->, thick, gray!70},
  skip/.style={->, thin, red!60},
  addnode/.style={circle, draw=gray!60, fill=white, inner sep=2pt, font=\tiny\bfseries}
]

\node[comp=violet, minimum width=1.1cm] (emb) {Embed};
\node[comp=violet, right=0.5cm of emb, minimum width=0.9cm] (pos) {+PE};

\node[draw=orange!60, fill=orange!5, rounded corners=3pt, 
      right=0.8cm of pos, minimum width=4.2cm, minimum height=2.6cm] (tblock) {};
\node[font=\tiny\bfseries, text=orange!70, above=0.1cm of tblock] {Transformer $\times N$};

\node[addnode] at ([xshift=-1.4cm, yshift=0.4cm]tblock.center) (add1) {+};
\node[comp=teal, minimum width=1.1cm, right=0.4cm of add1] (attn) {Self-Attn};
\node[comp=green, minimum width=0.8cm, right=0.4cm of attn] (ln1) {LN};

\node[addnode] at ([xshift=-1.4cm, yshift=-0.7cm]tblock.center) (add2) {+};
\node[comp=cyan, minimum width=0.9cm, right=0.4cm of add2] (ffn) {FFN};
\node[comp=green, minimum width=0.8cm, right=0.4cm of ffn] (ln2) {LN};

\node[comp=blue, right=0.8cm of tblock, minimum width=0.9cm] (proj) {Proj};
\node[comp=gray, right=0.5cm of proj, minimum width=1.0cm] (soft) {Softmax};

\draw[arr] (emb) -- (pos);
\draw[arr] (pos.east) -- ([xshift=-0.15cm]tblock.west) |- (add1.west);
\draw[arr] (add1) -- (attn);
\draw[arr] (attn) -- (ln1);
\draw[arr] (ln1.south) -- ++(0,-0.25) -| (add2.east);
\draw[arr] (add2) -- (ffn);
\draw[arr] (ffn) -- (ln2);
\draw[arr] (ln2.east) -- ++(0.15,0) |- ([yshift=0]tblock.east);
\draw[arr] (tblock.east) -- (proj.west);
\draw[arr] (proj) -- (soft);

\draw[skip, rounded corners=3pt] (pos.north) -- ++(0,0.9) -| (add1.north) node[pos=0.25, above, font=\tiny, text=red!70] {residual};
\draw[skip, rounded corners=3pt] (ln1.north) -- ++(0,0.35) -| ([xshift=0.15cm]add2.north east) node[pos=0.25, above, font=\tiny, text=red!70] {residual};

\node[font=\tiny, text=gray!60, below=0.15cm of emb] {tokens};
\node[font=\tiny, text=gray!60, below=0.15cm of soft] {output};

\end{tikzpicture}
}
\caption{GPT architecture: tokens are embedded with positional encoding, processed through $N$ transformer blocks (self-attention $\rightarrow$ layer norm $\rightarrow$ FFN $\rightarrow$ layer norm, with residual connections at each stage), then projected to output vocabulary.}
\Description{Block diagram of the GPT pipeline: token embeddings are combined with positional encoding and fed into a repeated transformer block. Inside the block, self-attention and a feed-forward network (FFN) are each followed by LayerNorm, with residual connections. Finally, the output is projected to the vocabulary and passed through a softmax to produce next-token probabilities.}
\label{fig:gpt_architecture}
\end{figure*}

As described in Figure~\ref{fig:gpt_architecture}, the GPT model architecture is designed to capture and generate human-like text by processing input through a series of transformer blocks.
Each block enhances the model's understanding of language context and structure, allowing for the generation of coherent and contextually relevant text.
This mechanism allows the model to simulate various forms of written content, including Fake News, by leveraging learned patterns from extensive data sets.

\section{Overview of Recent Developments}
\label{overview}

This section provides a condensed review of developments since our 2024 survey~\cite{loth_blessing_2024}, organized around key themes.

\subsection{The Agentic Shift in Misinformation}

The 2025--2026 literature reveals a shift in the threat landscape: the human actor is increasingly removed from the loop, replaced by autonomous agents.
Tseng et al. (2026) demonstrate multi-agent pipelines that operationalize the DISARM framework for investigating Foreign Information Manipulation and Interference (FIMI)~\cite{tseng2026agentic}.
While their work focuses on defensive applications, the architecture illustrates the dual-use potential: specialized agents can collaboratively map manipulative behaviors to standardized Tactics, Techniques, and Procedures (TTPs).

This implies that future detection systems must operate at the \emph{behavioral level}---identifying agent strategies---rather than the content level, as content will be hyper-optimized to evade static classifiers.
The ACM Europe Technology Policy Committee (2025) highlights this as a systemic risk, noting that current regulatory frameworks may govern ``models'' but fail to address emergent behaviors of ``agents'' exhibiting persistent operation and adaptive learning~\cite{acmetpc2025systemic}.

\subsection{Advances in LLM-Based Misinformation}

Sandrini and Somogyi (2023) analyze GenAI's effects on news consumption, finding that early-stage GenAI leads consumers toward deceptive content, though benefits emerge after reaching development thresholds~\cite{sandrini_generative_2023}.
Kumar et al. (2025) examine generative-AI-driven misinformation and propose counter-measures~\cite{kumar2025peeping}.

\subsection{Detection and Mitigation Strategies}

Detection approaches have evolved from rule-based systems to LLM-based methods.
Franco et al. (2025) propose using LLMs to prevent accidental sharing of misinformation~\cite{herder_preventing_2025}, while Sallami and Aïmeur (2025) review prevention techniques beyond detection~\cite{sallami_exploring_2025}.
Research has also revealed gender biases in detection systems~\cite{sallami_gender_2024}, prompting development of fairness frameworks~\cite{sallami_fairframe_2025}.
Tommasel and Rodriguez (2025) address identifying misinformation spreaders in social networks~\cite{tommasel_countering_2025}.

\subsection{Social Media and User Behavior}

Godoy (2024) examines the moral intuitions of fake news spreaders~\cite{godoy_moral_2024}, while Tintarev et al. (2024) study transparency in news recommendation systems~\cite{knijnenburg_transparency_2024}.
Herder and Staring (2024) analyze ``digital junkfood'' consumption patterns on social media~\cite{herder_junkfood_2024}.

Critical to policy discussions is the ``Transparency Penalty'' identified by Nakano et al. (2026): disclosing AI authorship generally erodes perceived trustworthiness, competence, and warmth~\cite{Nakano2026Perception}.
However, this effect is moderated by AI literacy---users with higher literacy are more tolerant of AI assistance and may even appreciate it.
This finding complicates simple policy prescriptions for mandatory labeling; if labels universally reduce trust regardless of content quality, they may contribute to the ``rational discounting'' of evidence predicted by Ferrara's Generative AI Paradox~\cite{Ferrara2026Paradox}.

\subsection{Deepfakes and Multimodal Misinformation}

Chun et al. (2024) find that older adults face challenges spotting deepfakes~\cite{chun2024can}, while Verma (2024) shows that a single deceptive video could affect geopolitical relations~\cite{verma2024one}.

The detection approach has shifted in 2026: single-modality detectors focusing on visual artifacts (warping, blending boundaries) or audio artifacts (robotic phrasing) are less effective against high-quality GenAI content.
The frontier is \emph{cross-modal consistency checking}.
Hussain et al. (2026) demonstrate that the most reliable signal is temporal inconsistency between modalities---specifically, subtle desynchronization between lip movements and speech audio, or semantic mismatch between visual context and audio narrative~\cite{Hussain2026Decoding}.
Their Synchronization-Aware Feature Fusion (SAFF) and Cross-Modal Graph Attention Networks (CM-GAN) architectures achieve 98.76\% accuracy on benchmarks like FaceForensics++ by explicitly modeling these cross-modal correlations.

Recent advances in Large Vision-Language Models (LVLMs) enable cross-modal semantic analysis~\cite{ai_paradigm_2026}, with multi-agent frameworks showing promise for complex verification tasks~\cite{shopnil_mirage_2025}.

\subsection{Ethical and Governance Considerations}

Aïmeur (2025) addresses privacy concerns in generative AI~\cite{aimeur_privacy_2025}, while Schrittwieser et al. (2024) examine trust and safety online~\cite{weippl_safe_2024}.
The EU AI Act and similar frameworks are shaping deployment requirements for high-risk AI applications.
Cena and Grasso (2025) explore users' mental models of conversational agents~\cite{cena_mental_2025}.

The 2025--2026 period marks the transition of C2PA from an emerging initiative to a global infrastructure standard.
C2PA Specification v2.3 (December 2025) introduced support for live video streaming---addressing a major gap in real-time news verification---and manifests for unstructured text, extending provenance beyond media files to LLM outputs~\cite{C2PASpec2025}.
Adoption has scaled significantly, with Google integrating C2PA Assurance Level 2 into Pixel camera hardware and TikTok implementing mandatory labeling for realistic AI content.
However, the Center for Democracy and Technology highlights a ``validity gap'': provenance proves origin (who signed it), not truth (is it factual?)~\cite{CDT2025Provenance}.
Privacy concerns regarding embedded metadata (location, author identity) remain particularly acute for activists and whistleblowers.

\section{Methodological Contributions: The Epistemic Security Experimental Pipeline}
\label{findings}

To operationalize the study of AI-mediated deception and quantify phenomena such as the ``Generative AI Paradox''~\cite{Ferrara2026Paradox} and the ``Agentic Shift,'' we introduce an experimental apparatus comprising two coupled components functioning as a closed-loop system.

\subsection{\ToolGen{}: Controlled Stimulus Engine}

\textit{\ToolGen}\footnote{\url{https://github.com/aloth/RogueGPT}} addresses the reproducibility challenge in misinformation research by replacing static datasets with a deterministic generation engine.
Through a formal configuration schema, it enables the controlled injection of generative variables---specifically Model Architecture ($M$), Temperature ($T$), Style ($S$), and Format ($F$)---into the experimental design.
This allows for the isolation of causal factors, represented formally as $\text{Stimulus} = f(M, T, S, F)$.

The engine ensures complete provenance by serializing the full generative context (system prompts, generation parameters) alongside the artifact, enabling retrospective analysis of deceptive strategies.
Integration with OpenAI and Azure OpenAI APIs supports multi-model comparison across LLM versions, while manual entry capability allows incorporation of human-written control stimuli.

\subsection{\ToolEval{}: Psychometric Evaluation Platform}

\textit{\ToolEval}\footnote{\url{https://github.com/aloth/JudgeGPT}}~\cite{loth2026eroding} serves as the measurement instrument for human epistemic resilience.
Unlike binary classification tasks common in detection research, \ToolEval{} employs continuous psychometric scales to capture the ambiguity of perception and the calibration of user confidence.
Participants rate perceived origin (definitely human to definitely machine-generated), perceived veracity (definitely legitimate to definitely fake), and topic familiarity on graded scales.

By integrating response latency metrics and demographic profiling, the platform facilitates intersectional analysis of susceptibility.
The architecture supports testing the efficacy of inoculation interventions~\cite{Spearing2025Countering}, measuring whether prebunking warnings effectively engage analytical processing in real-time consumption environments.

\subsection{Closed-Loop Verification System}

The integration of these components via a unified document-oriented data topology allows for measurement of the ``Perception-Accuracy Gap.''
Researchers first use \ToolGen{} to generate stimuli under controlled conditions, specifying model, style, format, and language parameters.
Generated fragments with full provenance metadata are stored in a shared MongoDB collection.
Participants access \ToolEval{}, which retrieves fragments and presents them for evaluation.
Responses are stored with links to fragment metadata, creating a dataset that enables precise attribution of perception effects to generation parameters.

This pipeline provides a standardized approach for measuring ``Deceptive Potential,'' allowing quantification of threat escalation as generative models evolve.

\subsection{Empirical Findings}

Our studies using this apparatus are reported in companion publications~\cite{loth2026eroding,loth2026verification}.
Data collection reveals that participants struggle to distinguish GPT-4 generated content from human-written text, with accuracy rates approaching chance levels for certain news styles~\cite{loth2026eroding,loth2026symposium}.
A perception-accuracy gap exists: increased suspicion does not improve detection accuracy, and asymmetric cognitive fatigue degrades fake detection by 10.2 percentage points under sustained exposure~\cite{loth2026killchaingi}.
Demographic predictors show weak effects: political orientation is only weakly associated with detection performance~\cite{loth2026eroding}, and age shows no significant correlation with accuracy, challenging the ``digital native'' assumption~\cite{loth2026symposium}.
Familiarity with fake news is associated with higher suspicion but only negligibly with detection accuracy~\cite{loth2026killchaingi}.

These findings align with broader research indicating that LLM-generated content is increasingly difficult to detect~\cite{maronikolakis_identifying_2021,goh2024humans}.

\section{Synthesis and Mitigation Strategies}
\label{state}

Simon et al. (2023) discuss ethical considerations for AI in journalism, emphasizing the balance between benefits and risks~\cite{simon_misinformation_2023}.
Weisz et al. (2023) propose design principles for generative AI that prioritize safety~\cite{weisz_toward_2023}.

\subsection{Mitigation Approaches}

Countering Generative AI's adverse effects requires strategies spanning technology, education, and policy:

\textbf{Technological Approaches:} Detection algorithms using the same LLMs employed for generation have demonstrated measurable effectiveness~\cite{sallami_deception_2024,herder_preventing_2025}.
However, the competition between generation and detection has become adversarial.
Tahmasebi et al. (2026) demonstrate that many state-of-the-art detectors rely heavily on sentiment correlations---assuming, for instance, that fake news is negative or inflammatory---making them vulnerable to ``sentiment attacks'' that rewrite false claims to sound neutral or positive~\cite{Tahmasebi2026AdSent}.
Their AdSent framework shows that such attacks can degrade detection performance (F1-score) by over 20\%, confirming that adversaries are now optimizing latent features to traverse detector decision boundaries.
This necessitates sentiment-agnostic training strategies that force models to learn veracity features independent of emotional tone.

Recent work on multimodal LLM-based detection systems, such as TRUST-VL~\cite{yan_trustvl_2025} and agentic multi-persona frameworks~\cite{bukke_agentic_2025}, demonstrates improved accuracy through combining multiple reasoning perspectives.
Tool-augmented agents using Monte Carlo Tree Search have achieved high performance in complex multimodal verification~\cite{cui_t2agent_2026}.
Future detectors should be adversarially aware and sentiment-agnostic, moving beyond stylistic analysis toward features that capture veracity rather than surface patterns.

\textbf{Inoculation and Prebunking:} Lewandowsky and Van Der Linden's (2021) inoculation theory has gained traction, emphasizing preemptive education to build resilience~\cite{lewandowsky_countering_2021}.
Spearing et al. (2025) provide empirical support in the GenAI context: ``pre-emptive source discreditation''---warning users about the manipulative tactics of a source before exposure---is more effective than reactive debunking~\cite{Spearing2025Countering}.
This is relevant for GenAI content, where the volume makes reactive fact-checking impractical.
Our \ToolEval{} platform offers opportunities to measure not just detection accuracy but also the efficacy of such inoculation interventions.

\textbf{Provenance and Authenticity Infrastructure:} Content authenticity initiatives provide cryptographic verification of content origin as an alternative to detection-based approaches.
The C2PA standard has matured with v2.3 supporting live streaming and text manifests~\cite{C2PASpec2025}, while Google DeepMind's open-source SynthID Text provides a complementary watermarking layer using tournament-based token probability adjustment that resists modification yet remains invisible to humans~\cite{dathathri2024scalable}.
This ``defense-in-depth'' approach---if metadata is stripped, the watermark may remain---aligns with our Origin Lens framework, which performs privacy-first on-device C2PA verification, combining cryptographic provenance with heuristic metadata analysis, watermark detection, and graded confidence indicators~\cite{loth2026originlens}.
However, the ``validity gap'' remains: provenance proves origin, not truth~\cite{CDT2025Provenance}.
Challenges persist around manifest stripping, analog-hole attacks, and privacy implications for whistleblowers.

\textbf{Platform Design:} Herder and Staring explore interface designs that help users manage social media consumption and avoid misinformation~\cite{herder_junkfood_2024}.
Transparency mechanisms and friction-inducing interventions can slow the reflexive sharing that accelerates misinformation spread.

\textbf{Collaborative Measures:} Shu et al. (2020) explore collaborative approaches involving governments, private sector, and civil society~\cite{shu_combating_2020}.
The development of shared benchmarks and evaluation frameworks enables progress in detection capabilities.

\subsection{Stakeholder Landscape}

Key stakeholders include academics developing detection algorithms~\cite{godoy_moral_2024}, technology developers deploying AI detection systems, platforms enforcing misuse prevention, and policymakers creating regulations~\cite{aimeur_privacy_2025}.

\subsection{Open Research Directions}

Several unresolved issues demand continued attention.
\emph{Adversarial robustness} remains a challenge: the arms race has moved to the feature level, with adversaries optimizing latent features to evade detectors, as demonstrated by sentiment attacks that degrade F1-scores by over 20\%~\cite{Tahmasebi2026AdSent,cocchi_unveiling_2023}.
\emph{Multimodal challenges} require detection approaches that can analyze cross-modal semantic consistency and identify out-of-context manipulations~\cite{ai_paradigm_2026,yan_trustvl_2025}.
Global misinformation campaigns necessitate \emph{cross-lingual detection} capabilities~\cite{schutz_ait_fhstp_2022,li_multicaption_2026}, while ensuring \emph{bias and fairness} in detection systems requires explicit attention to avoid creating new forms of harm~\cite{sallami_fairframe_2025}.
The operationalization of Foreign Information Manipulation and Interference (FIMI) through multi-agent pipelines demands \emph{behavioral-level detection} that analyzes Tactics, Techniques, and Procedures (TTPs) within frameworks like DISARM rather than isolated artifacts~\cite{tseng2026agentic,ferrara_genai_2024,bukke_agentic_2025}.
Finally, \emph{digital ecosystem resilience} requires infrastructure-level approaches to information integrity, including provenance standards and platform design interventions~\cite{loth2026originlens,fastowski_knowledge_drift_2024}.

Future research must navigate technological innovations alongside broader societal, ethical, and psychological dimensions of this challenge.

\section{Conclusion}
\label{conclusion}

This paper has examined how Generative AI has changed the disinformation landscape since our 2024 survey.
We documented developments in LLM capabilities, multimodal misinformation, and the dual-use nature of these technologies for both generation and detection.

Our methodological contributions---\textit{\ToolEval}~\cite{loth2026eroding} and \textit{\ToolGen}---provide an experimental pipeline for studying human perception of AI-generated news.
Our companion studies using this pipeline reveal that participants struggle to distinguish LLM-generated content from human-written text, with accuracy approaching chance levels for certain news styles~\cite{loth2026eroding,loth2026symposium}.
Our longitudinal expert survey found that specialists view large-scale text generation as posing systemic risks of ``epistemic fragmentation'' and ``synthetic consensus''---findings now formalized in Ferrara's ``Generative AI Paradox,'' which argues that the cost of verification has become prohibitively high compared to the cost of generation, rendering trust economically irrational~\cite{Ferrara2026Paradox}.
Experts express skepticism toward purely technical detection tools, preferring provenance standards aligned with emerging ``epistemic security'' objectives~\cite{loth2026verification,IPIE2025Epistemic}.

Several key insights emerge from this analysis.
First, the threat has evolved beyond ``fake news'' to ``Synthetic Reality''---a layered stack comprising synthetic content, identity, interaction, and in\-sti\-tu\-tions---requiring defenses that address each layer~\cite{Ferrara2026Paradox}.
The dual-use nature of LLMs offers opportunities for ``fighting fire with fire'' approaches~\cite{sallami_deception_2024,herder_preventing_2025}, while multimodal misinformation requires detection approaches that analyze cross-modal semantic consistency~\cite{ai_paradigm_2026,yan_trustvl_2025}.
Prevention and prebunking strategies may prove more effective than reactive detection, as inoculation theory gains empirical support~\cite{lewandowsky_countering_2021,sallami_exploring_2025}.
Bias and fairness in detection systems require explicit attention to avoid creating new forms of harm~\cite{sallami_fairframe_2025}.
Digital ecosystem resilience requires infrastructure-level interventions including content provenance standards and platform design changes aligned with epistemic security objectives~\cite{loth2026originlens,IPIE2025Epistemic}.
Finally, the rise of agentic AI systems introduces new vectors for scaled misinformation campaigns that demand behavioral-level detection and proactive governance~\cite{tseng2026agentic,acmetpc2025systemic}.

Our review suggests that purely technical countermeasures---such as watermarking or detection classifiers---face significant challenges due to the rapid adaptability of generative models.
The competition between generation and detection capabilities continues, with current mitigation strategies struggling to keep pace.

Future research should explore proactive approaches including adversarial testing, provenance infrastructure, and governance frameworks.
Our \ToolEval{}-\ToolGen{} pipeline offers one foundation for investigating human perception of AI-generated content.

Addressing the adverse impacts of generative AI on information quality will require efforts combining technical safeguards, media literacy initiatives, platform accountability, and policy frameworks.

As our study continues, we invite experts to participate in our ongoing survey: \url{https://github.com/aloth/verification-crisis}.

\bibliographystyle{ACM-Reference-Format}
\balance
\bibliography{main}

\end{document}